\documentstyle[12pt,preprint,aps,tighten,epsfig,floats]{revtex}

\def\beq{\begin{equation}}
\def\eeq{\end{equation}}
\def\beqa{\begin{eqnarray}}
\def\eeqa{\end{eqnarray}}
\def\be{\begin{equation}}
\def\ee{\end{equation}}
\def\bea{\begin{eqnarray}}
\def\eea{\end{eqnarray}}
\begin{document}

\preprint{
\vbox{\hbox{UWThPh-2001-39}}}
\title{Comment on
`Analysis of ${\cal O}(p^2)$ Corrections to
$\langle \pi \pi | {\cal Q}_{7,8} | K \rangle$'}
\author{Vincenzo Cirigliano}
\address{Inst. f\"ur Theoretische Physik,
University of Vienna \\
Boltzmanngasse 5, Vienna A-1090 Austria \\
vincenzo@thp.univie.ac.at}
\author{Eugene Golowich}
\address{Physics Department, University of Massachusetts\\
Amherst, MA 01003 USA \\
golowich@physics.umass.edu \\}
\maketitle
\thispagestyle{empty}
\setcounter{page}{0}
\begin{abstract}
\noindent We extend in several respects our earlier work on ${\cal O}(p^2)$ 
corrections to matrix elements of the electroweak penguin operator 
${\cal O}_{\rm ewp}$.  First, to facilitate comparison with certain 
lattice studies we calculate ${\cal O}(p^2)$ corrections to 
$\langle \pi | {\cal O}_{\rm ewp} | K \rangle$ in the SU(3) limit 
of equal light quark masses.  Next, we demonstrate how an apparent 
disagreement in the literature regarding whether higher order chiral 
contributions increase or decrease 
$\langle (\pi \pi)_{{\rm I}=2}|{\cal O}_{\rm ewp}| K \rangle$ 
is simply a consequence of how the leading order chiral amplitude 
is defined.  Finally, we address an aspect of the $\epsilon'/
\epsilon$ problem by estimating ${\cal O}(p^2)$ corrections 
to recent determinations of 
$\langle (\pi \pi)_{{\rm I}=2}|{\cal Q}_{7,8} | K \rangle$ which 
were carried out in the chiral limit.
\end{abstract}

\vspace{1.0in}

\section{Introduction}
There is currently great interest in matrix elements of 
the four-quark operators ${\cal Q}_{7,8}$, both  
in the phenomenology of $\epsilon'/\epsilon$ and in 
lattice studies.  Not surprisingly, chiral perturbation 
theory (ChPT) provides an important theoretical context 
for progress in this area.  At chiral order ${\cal O} (p^0)$, 
both ${\cal Q}_7$ and ${\cal Q}_8$ are represented uniquely by the 
electroweak penguin operator ${\cal O}_{\rm ewp}$.  
In a recent paper~\cite{cg}, we performed a ChPT 
analysis of one-loop corrections to $K \to \pi$ and 
$K \to 2 \pi$ matrix elements of ${\cal O}_{\rm ewp}$.\footnote{In 
Ref.~\cite{cg} the following substitutions should be made to 
correct typographical errors:
\begin{enumerate}
\item Replace $F_0$ by $F$ in Eq.~(13).
\item Omit $F^2$ from Eq.~(20).
\item In $I_{K^{+} \pi^{+}} (q^2)$ of Eq.~(29), replace
$\dots + {\bar A} (m_{K}^{2}) \dots$ by
$\dots - {\bar A} (m_{K}^{2}) \dots$.
\item Divide the final line of Eq.~(29) by two.
\end{enumerate}
These errors occurred entirely in the printing process;
none of the results of the paper are changed.}  
The purpose of this paper is to expand upon several aspects 
of Ref.~\cite{cg}.

First, the calculation in Ref.~\cite{cg} employed
physical values for the meson masses $m_\pi$, $m_K$ and
$m_\eta$. This turns out to be rather general compared to what
some current lattice calculations need~\cite{soni}.  Work on
kaon-to-pion matrix elements done with domain wall quarks
has been performed in the SU(3) symmetric limit~\cite{bs,cp-pacs}.
In fact, reference to the SU(3) limit has been a common strategy
in certain lattice simulations for quite some time~\cite{bdspw}.
Below, we shall report the (nontrivial) restriction of our $K$-to-$\pi$ 
matrix elements to the equal quark mass case of $m_u = m_d = m_s$.
In this work we present results valid in the case of unquenched QCD. 
Analogous  results in the quenched and partially quenched case 
can be found in Ref.~\cite{gp01}.

Another feature of Ref.~\cite{cg} was the determination of the
fractional shift $\Delta_2$ for the ${\cal O}(p^2)$
corrections to the chiral limit determination of
$\langle (\pi\pi)_{I=2}| {\cal Q}_{7,8}
| K^0 \rangle$.  In particular, we
discussed why the sign for $\Delta_2$ is
opposite to that expected from unitarization
approaches ({\it e.g.} see Ref.~\cite{pp}) based
on the Omn\`es equation. Results in Ref.~\cite{pps}
would appear to contradict this finding.  It turns out,
however, that Ref.~\cite{cg} and Ref.~\cite{pps}
normalize the so-called `chiral limit result'
({\it i.e.} the leading order term in a chiral perturbation
theory expansion) in different ways.  In order to eliminate
any undue confusion in future literature that this issue might
cause, we carefully identify the source of the difference.

Finally, using the chiral limit normalization
for $\langle (\pi\pi)_{I=2}| {\cal Q}_{7,8}
| K^0 \rangle$ appearing in
Refs.~\cite{kpr,bgp,cdgm}, we discuss the ${\cal O} (p^2)$
corrections to such determinations.
Knowing the size of such corrections is important
in order to compare the predictions of Refs.~\cite{kpr,bgp,cdgm}
with recent lattice QCD determinations~\cite{lattice}.

\section{Analysis}
In this section, we shall be concerned with both $K$-to-$\pi$ and 
$K$-to-$2\pi$ matrix elements of ${\cal O}_{\rm ewp}$. 
We recall the ChPT definition 
${\cal O}_{\rm ewp} \equiv g \,
 \mbox{Tr} \, \left[ \lambda_{6} U Q U^{\dagger} \right]$ where 
$Q = ~diag~(2/3,-1/3,-1/3)$ is the quark charge matrix 
and $U \equiv \exp(i \lambda_k \Phi_k /F)$ 
is the matrix of light pseudoscalar fields.  We denote 
the pseudoscalar meson decay constant in lowest order by $F$. 

\subsection{The SU(3) Limit of $\langle \pi | {\cal O}_{\rm ewp} 
| K \rangle$}
Let us denote any amplitude evaluated in the SU(3) limit
with a superbar (${\overline {\cal M}}_i$) and likewise
for the meson masses,
\beq
{\overline m}^2 \ = \ m_\pi^2 \ = \ m_K^2 \ = \ m_\eta^2 \qquad
{\rm (SU(3) \ limit)} \ \ .
\label{mesmass}
\eeq
The ${\cal O}(p^0)$ amplitudes are unaffected by passage to the
SU(3) world,
\beq
{\overline{\cal M}}_{K^{+} \to \pi^{+}}^{(0)} =
{\cal M}_{K^{+} \to \pi^{+}}^{(0)} = { 2 g \over F^2}
\ , \qquad
{\overline {\cal M}}_{K^{0}\to \pi^{0}}^{(0)} =
{\cal M}_{K^{0}\to \pi^{0}}^{(0)} = 0
\label{finite1.5}
\eeq
and
\beq
-i {\overline{\cal M}}_{K^0 \to \pi^+\pi^-}^{(0)} =
-i {\cal M}_{K^0 \to \pi^+\pi^-}^{(0)} = -
 {\sqrt{2} g \over F^3}  \ , \qquad
-i {\overline{\cal M}}_{K^0\to \pi^0 \pi^0}^{(0)} =
-i {\cal M}_{K^0\to \pi^0 \pi^0}^{(0)} = 0  \ \ .
\label{finite5.5}
\eeq

It is for the ${\cal O}(p^2)$ amplitudes that the SU(3)
limit is nontrivial.  Calculation reveals the full next-to leading 
order amplitudes to be 
\beqa 
\overline{\cal M}_{K^+ \to \pi^+}^{(0+2)} &=&
\frac{2 \, g}{F^2} \left[ 1 \, - \,  
{{3 \, \overline m}^2 \over (4 \pi F)^2} \log \frac{{\overline m}^2}{
\mu_\chi^2} \, + \, {{\overline m}^2 \over F^2}  \, \bigg[ 
K_{++}^r (\mu_\chi) - 24 L_4^r (\mu_\chi) - 8 L_5^r (\mu_\chi) \bigg] 
\right] \nonumber \\
&=& \frac{2 \, g}{F_\pi F_K} \left[ 1 \, - \, {{6 \, 
\overline m}^2 \over (4 \pi F)^2} \log \frac{{\overline m}^2}{\mu_\chi^2}
\, + \, {{\overline m}^2 \over F^2} \,  K_{++}^r (\mu_\chi)  
\right]  \ , \label{su3} \\
\overline{\cal M}_{K^0 \to \pi^0}^{(0+2)} &=&
\frac{\sqrt{2} \, g}{F^2} \left[  - \,  {{ \, \overline m}^2 
\over (4 \pi F)^2} \left( 1 + \log \frac{{\overline m}^2}{
\mu_\chi^2} \right) \, + \, {{\overline m}^2 \over F^2}  \,
K_{00}^r (\mu_\chi) \right] \ \ .  \nonumber 
\eeqa
In the above, $\mu_\chi$ is an arbitrary energy scale, 
$L_{4,5}^r (\mu_\chi)$ 
are finite, scale-dependent low energy constants (LECs) of the 
${\cal O}(p^4)$ strong chiral lagrangian~\cite{gl851} and 
$K_{++,00}^r (\mu_\chi)$ are finite, scale-dependent combinations 
of the counterterms ($\{c_i\}$) defined in Ref.~\cite{cg}, 
\beqa
K_{++} &=& {2 \, F^2 \over g} \, \bigg[
 {1 \over 3} \left(  c_1 - c_3 \right)
+ {7 \over 3} c_4 + 2 c_5 + 3 c_6 \bigg] \ ,
\nonumber \\
K_{00} &=& {  F^2 \over g} \, 
\bigg[ {1 \over 3} c_1 + c_2 + { 2\over 3} c_3
- {2 \over 3} c_4 \bigg] \ \ .
\eeqa
Using the results of Ref.~\cite{cg} one can verify that 
the combinations $K_{++,00}^r (\mu_\chi)$ compensate for the 
explicit scale dependence of the chiral logarithms and 
the implicit scale dependence of the finite LECs $L^r_{4,5}$. 
The presence of $L_{4,5}$ can be understood as affecting 
the ${K^+ \to \pi^+}$ transition via wavefunction renormalization. 
We note that the two forms for $\overline{\cal M}_{K^+ \to 
\pi^+}^{(0+2)}$ displayed in Eq.~(\ref{su3})  
correspond to the choice of either keeping $L_{4,5}$
explicit or absorbing them in the renormalization of $F_\pi, F_K$
(as done in Ref.~\cite{cg}). Recall that in the SU(3) limit 
the relation between $F_\pi, F_K$ and $F$ is given by~\cite{gl851}  
\beq
F_{\pi,K} = F \, \left[ 1  - {3 \over 2} \,
{{\overline m}^2 \over (4 \pi F)^2} \log \frac{{\overline m}^2}{\mu_\chi^2}\,
+ \, {{\overline m}^2 \over F^2} 
\bigg( 12 \, L_4^r (\mu_\chi) \, + \, 4 \, L_5^r (\mu_\chi)  \bigg) 
\right] \ . 
\eeq

\subsection{Alternative Definitions of `The Leading Chiral Term'}

For the remainder of this paper, we leave the SU(3) 
limit and hereafter employ physical values for all particle masses. 
Consider K-to-$2\pi$ matrix elements of the operators
${\cal Q}_{7,8}$ written as
\beq
{\cal M}_{I} \equiv \langle (\pi \pi)_I  |
{\cal Q}_{7,8} | K \rangle
= {\cal M}_{I}^{(0)} \cdot
\left( 1 + \Delta_I \right) \qquad (I = 0,2) \ \ ,
\label{chiral1}
\eeq
where ${\cal M}_{I}^{(0)}$ is evaluated in the
chiral world and $\Delta_I$ gives the
fractional ${\cal O}(p^2)$ correction.
In particular we found in Ref.~\cite{cg}
for the isospin I=2 case that
chiral corrections {\it increase}
the chiral limit value by about $27\%$
($\Delta_2^{\rm CG} = + 0.27 \pm 0.27$), in seeming contrast with
the recent claim~\cite{pps} that
chiral loops {\it reduce} the chiral limit value by about $50\%$.

We wish to explain the origin of this discrepancy.
It is not due to mistakes in either
Ref.~\cite{cg} or Ref.~\cite{pps} but rather to the fact that
the chiral limit result is normalized
differently in these two papers. 
Our first observation is that in Ref.~\cite{cg} we work with
a dimensionful coupling $g$ (of dimension six in mass),
while in Ref.~\cite{pps} a dimensionless coupling is used
(we denote it here by $\overline{g}$),
\beq
{\cal O}_{\rm ewp}  \ = \
\left\{
\begin{array}{ll}
g \  \langle \lambda_6 \,  U \, Q\, U^\dagger \rangle
& ({\rm Ref.}~\cite{cg})
\nonumber  \\
F^6 \, \overline{g} \  \langle \lambda_6 \,  U \, Q\, U^\dagger \rangle
& ({\rm Ref.}~\cite{pps}) \ \ ,
\end{array}
\right.
\label{compare1}
\eeq
implying the leading order matrix elements 
\beq
i {\cal M}_{2}^{(0)} \ = \
\left\{
\begin{array}{ll}
2g/(3 F^3) 
& \qquad ({\rm Ref.}~\cite{cg})
\nonumber  \\
2 F^3 \overline{g}/3 
& \qquad ({\rm Ref.}~\cite{pps}) \ \ .
\end{array}
\right.
\label{compare2}
\eeq
Moreover, the chiral loop corrections are defined 
in the two references as 
\beq
i {\cal M}_{2}^{(0+2)} \ = \
\left\{
\begin{array}{ll}
\displaystyle{2\over 3} {g \over F_\pi^2 F_K} \, \left( 1 + 
\Delta_2^{\rm CG} \right)
& ({\rm Ref.}~\cite{cg})
\nonumber  \\
\displaystyle{2 \over 3} \, F_\pi^3 \,  \overline{g}  
\, \left( 1 + \Delta_2^{\rm PPS} \right)
& ({\rm Ref.}~\cite{pps}) \ \ .
\end{array}
\right.
\label{compare3}
\eeq
That is, both analyses shift some one-loop
terms (the ratios $F/F_\pi$ and $F/F_K$)
into the definition of the leading order matrix element.
Clearly this makes no difference at all if one sums the
leading and next-to-leading terms. However, this will affect
what the two references call the
`next-to-leading term' ($\Delta_2^{\rm CG}$ vs 
$\Delta_2^{\rm PPS}$) and explains the
difference in their stated `chiral corrections'.\footnote{For a 
toy version of this point, if ChPT analyses of researchers 
A,B are written as ${\cal M}_A = 10 - 2$ and ${\cal M}_B = 5 + 3$, 
then the percentage corrections will be very different, 
$\Delta_A = - 20\%$ and $\Delta_B = + 60\%$.} 
The large difference in the stated results is accounted for by
the large powers of  $F/F_\pi$ and $F/F_K$ needed to relate
$\Delta_2^{\rm CG}$ to $\Delta_2^{\rm PPS}$.
Moreover, neither of the two definitions coincides with the one 
given in Eq.~(\ref{chiral1}), where $\Delta_I$ includes {\it all} 
the corrections of ${\cal O}(p^2)$. 

\subsection{Estimate of Pure Next-to-Leading Order Corrections}
Some recent papers~\cite{kpr,bgp,cdgm} are devoted to evaluating 
the $K \rightarrow \pi \pi$
electroweak penguin matrix elements in the chiral limit. The procedure
used there is to relate the dimensionful constant $g$ to
vacuum expectation values of appropriate dimension six operators.
The $K \rightarrow \pi \pi$ matrix elements are then obtained
by normalizing with the appropriate numerical factors and $1/F^3$,
corresponding to the first line in Eq.~(\ref{compare2}). 
The chiral corrections to these determinations (see also 
Ref.~\cite{bf}) are therefore given by 
$\Delta_2$ of Eq.~(\ref{chiral1}) and upon adopting the convenient 
reference scale as the $\rho$-meson mass ($\mu_\chi = m_\rho$) we find 
\beq
\Delta_2 = - 0.118 - 0.727 \, \frac{L_4^r (m_\rho)}{10^{-3}}
- 0.134 \, \frac{L_5^r (m_\rho)}{10^{-3}} + \Delta_2^{\rm (ct)} 
(m_\rho) \ \ ,
\label{result1}
\eeq
where $\Delta_2^{\rm (ct)}(m_\rho)$ is the contribution from the finite 
${\cal O}(p^2)$ electroweak counterterms, 
\beq
\Delta_2^{\rm (ct)} (m_\rho) = 
- {1 \over g} \, \bigg[ m_K^2 \left(  c_2^r + c_3^r 
- 2 c_4^r - 2 c_5^r - 4 c_6^r \right) - m_\pi^2 \left( 
c_1^r + c_2^r + 4 c_4^r + 4 c_5^r + 2 c_6^r \right) 
\bigg]_{\mu_\chi = m_\rho} \ \ .
\label{ewpct}
\eeq
In Eq.~(\ref{result1}), the first numerical factor comes 
from chiral loops evaluated at scale $m_\rho$ and the LECs $L_{4,5}^r$ 
(which enter via wavefunction renormalization) are normalized 
to $10^{-3}$. Analogous to the procedure adopted in Ref.~\cite{cg}, we 
can estimate the size of $\Delta_2^{\rm (ct)} (m_\rho)$ by varying 
the scale of the pure chiral loop term between 0.6 GeV and 1 GeV.  
This procedure yields $| \Delta_2^{\rm (ct)} (m_\rho) | \leq 0.20 $.  
We then use the value $L_5^r (m_\rho) = (1.4 \pm 0.5) 
\cdot 10^{-3}$ to arrive at the conservative estimate 
\beq
\Delta_2 = - \left( 0.30  \pm 0.21  + 0.727 \, \frac{L_4^r
(m_\rho)  }{10^{-3}}
\right) \ .
\label{result2}
\eeq
Since $L^r_4$ is poorly known, it is not possible 
to estimate $\Delta_2$ more precisely.  We note that 
$\Delta_2^{\rm CG,PPS}$ can be related to $\Delta_2$ by using 
the appropriate expressions for $F_\pi / F$ and $F_K / F$, and we 
have explicitly checked the agreement of 
Ref.~\cite{cg} and Ref.~\cite{pps} on this point.

Finally, in the large $N_c$ limit one has $L_4 =0$ and an explicit
expression of $\Delta_2^{\rm (ct)} $ in terms of $L_5^r$~\cite{pps}. 
This term is seen to cancel almost exactly the $L_5$ contribution 
from wavefunction renormalization~\cite{pps}, and one 
obtains\footnote{The
large $N_c$ estimate reported here only refers to the operator 
${\cal Q}_8$, which is of considerable phenomenological interest.} 
$\Delta_2^{N_c \rightarrow \infty} = - 0.08$.

\section{Summary of Results} 
We enumerate our conclusions as follows:
\begin{enumerate}
\item We have displayed in Eq.~(\ref{su3}) the K-to-$\pi$ matrix 
elements of the electroweak penguin operator ${\cal O}_{\rm ewp}$ 
in the limit of SU(3) flavor symmetry.  This will allow comparison 
with lattice studies which work in this kinematic regime.
\item We have pointed out how different definitions of 
`leading chiral order' amplitudes ({\it viz} Eq.~(\ref{compare3})) 
can lead to numerically 
distinct `chiral corrections', which differ not only in magnitude 
but even in sign.  Although such distinctions are a matter of chiral 
bookkeeping and have no intrinsic meaning, one must be careful 
to avoid misinterpretation.
\item We have provided in Eq.~(\ref{result2}) a numerical 
estimate of ${\cal O}(p^2)$ corrections to 
$\langle (\pi \pi)_{{\rm I}=2}|{\cal Q}_{7,8} | K \rangle$.   
Such corrections would modify recent chiral determinations 
of this matrix element.
Our results indicate that even in the extreme scenario of
$\Delta_2 \simeq - 0.50$ allowed by our uncertainties, the 
results of Refs.~\cite{kpr,cdgm} for 
$\langle (\pi\pi)_{I=2}| {\cal Q}_{8}|
K^0 \rangle$ remain significantly larger than cited lattice 
values~\cite{lattice}.
\end{enumerate}

\acknowledgements
The authors thank A. Soni for stimulating correspondence, 
A. Pich and J. Donoghue for useful discussions and 
X. Prades for a helpful suggestion.
This work was supported in part by the National
Science Foundation under Grant PHY-9801875.
The work of V.C. is funded by TMR, EC-Contract  
No. ERBFMRX-CT980169 (EURODA$\Phi$NE).

\eject


\begin{thebibliography}{99}

%\cite{Cirigliano:2000pv}
\bibitem{cg}
V.~Cirigliano and E.~Golowich,
``Analysis of ${\cal O}(p^2)$ corrections to $\langle \pi \pi | Q_{7,8} | K 
\rangle $,''
Phys.\ Lett.\ B {\bf 475}, 351 (2000)
[hep-ph/9912513].
%%CITATION = HEP-PH 9912513;%%

\bibitem{soni} A. Soni, private communication.

%\cite{Blum:1997jf}
\bibitem{bs}
T.~Blum and A.~Soni,
``QCD with domain wall quarks,''
Phys.\ Rev.\ D {\bf 56}, 174 (1997)
[hep-lat/9611030].
%%CITATION = HEP-LAT 9611030;%%

%\cite{AliKhan:2001bx}
\bibitem{cp-pacs}
A.~Ali Khan {\it et al.}  [CP-PACS Collaboration],
``Calculation of K $\to \pi$ matrix elements in quenched 
domain-wall QCD,''
Nucl.\ Phys.\ Proc.\ Suppl.\  {\bf 94}, 283 (2001)
[hep-lat/0011007].

%\cite{Bernard:1985wf}
\bibitem{bdspw}
C.~Bernard, T.~Draper, A.~Soni, H.~D.~Politzer and M.~B.~Wise,
``Application of chiral perturbation theory to $K \to 2 \pi$ decays,''
Phys.\ Rev.\ D {\bf 32}, 2343 (1985).
%%CITATION = PHRVA,D32,2343;%%

%\cite{Golterman:2001qj}
%\bibitem{Golterman:2001qj}
\bibitem{gp01}
M.~Golterman and E.~Pallante,
``Effects of quenching and partial quenching on penguin matrix elements,''
hep-lat/0108010.
%%CITATION = HEP-LAT 0108010;%%

%\cite{Pallante:2000qf}
\bibitem{pp}
E.~Pallante and A.~Pich,
``Strong enhancement of $\epsilon'/\epsilon$ through final state
interactions,''
Phys.\ Rev.\ Lett.\  {\bf 84}, 2568 (2000)
[hep-ph/9911233].
%%CITATION = HEP-PH 9911233;%%

%\cite{Pallante:2001he}
\bibitem{pps}
E.~Pallante, A.~Pich and I.~Scimemi,
``The standard model prediction for $\epsilon'/\epsilon$,''
hep-ph/0105011.
%%CITATION = HEP-PH 0105011;%%

\bibitem{kpr}
%\bibitem{Knecht:2001bc}
M.~Knecht, S.~Peris and E.~de Rafael,
``A critical reassessment of ${\cal Q}_7$ and ${\cal Q}_8$ 
matrix elements,''
Phys.\ Lett.\ B {\bf 508}, 117 (2001)
[hep-ph/0102017].
%%CITATION = HEP-PH 0102017;%%

\bibitem{bgp}
%\bibitem{Bijnens:2001ps}
J.~Bijnens, E.~Gamiz and J.~Prades,
``Matching the electroweak penguins ${\cal Q}_7$, 
${\cal Q}_8$ and spectral correlators,''
hep-ph/0108240.
%%CITATION = HEP-PH 0108240;%%

\bibitem{cdgm}
%\bibitem{Cirigliano:2001qw}
V.~Cirigliano, J.~F.~Donoghue, E.~Golowich and K.~Maltman,
``Determination of $ \langle(2 \pi)_{I=2}|Q_{7,8}|K^0\rangle $ 
in the Chiral Limit,''
hep-ph/0109113.
%%CITATION = HEP-PH 0109113;

\bibitem{lattice}
G. Martinelli, Talk given at 19th International Symposium on Lattice 
Field Theory (LATTICE 2001), Berlin, Germany, August 19-24, 2001.  

%\cite{Antonelli:1996gw}
\bibitem{bf}
V.~Antonelli, S.~Bertolini, M.~Fabbrichesi and E.~I.~Lashin,
``The $\Delta I = 1/2$ Selection Rule,''
Nucl.\ Phys.\ B {\bf 469}, 181 (1996) 
[hep-ph/9511341]; S.~Bertolini, J.~O.~Eeg and M.~Fabbrichesi,
``A New Estimate of $\varepsilon '/\varepsilon$,''
Nucl.\ Phys.\ B {\bf 476}, 225 (1996) [hep-ph/9512356]; 
S.~Bertolini, M.~Fabbrichesi and J.~O.~Eeg,
``Theory of the CP-violating parameter epsilon'/epsilon,''
Rev.\ Mod.\ Phys.\  {\bf 72}, 65 (2000) [hep-ph/9802405].

\bibitem{gl851}
%\bibitem{Gasser:1985gg}
J.~Gasser and H.~Leutwyler,
``Chiral perturbation theory: expansions in the mass of the strange quark,''
Nucl.\ Phys.\ B {\bf 250}, 465 (1985).
%%CITATION = NUPHA,B250,465;%%

\end{thebibliography}
\end{document}